\documentstyle[12pt,epsfig,amsmath,amssymb,natbib]{article}
\begin{document}

\title{\bf On the Ollivier-Poulin-Zurek definition of objectivity}

\author{{Chris Fields}\\ \\
{\it 815 E. Palace \# 14}\\
{\it Santa Fe, NM 87501 USA}\\ \\
{fieldsres@gmail.com}}
\maketitle

\begin{abstract}
The Ollivier-Poulin-Zurek definition of objectivity provides a philosophical basis for the environment as witness formulation of decoherence theory and hence for quantum Darwinism.  It is shown that no account of the reference of the key terms in this definition can be given that does not render the definition inapplicable within quantum theory.  It is argued that this is not the fault of the language used, but of the assumption that the laws of physics are independent of Hilbert-space decomposition.  All evidence suggests that this latter assumption is true.  If it is, decoherence cannot explain the emergence of classicality.  
\end{abstract}


\section{Introduction}
\label{Introduction}

In papers published in \textit{Physical Review Letters} and \textit{Physical Review} in 2004 and 2005 respectively, H. Ollivier, D. Poulin and W. H. Zurek proposed an operational definition of \textit{objectivity} for physical systems:

\begin{quotation}
``A property of a physical system is \textit{objective} when it is:
\begin{list}{\leftmargin=2em}
\item
1. simultaneously accessible to many observers,
\item
2. who are able to find out what it is without prior knowledge about the system of interest, and 
\item
3. who can arrive at a consensus about it without prior agreement.''
\end{list}
\begin{flushright}
\cite{zurek:04} p. 1; \cite{zurek:05} p. 3
\end{flushright}
\end{quotation}
This ``OPZ'' definition makes explicit the sense of ``objectivity'' being sought in the attempt to explain the ``emergence of classicality'' within decoherence theory.  It is stated within a theoretical context that treats minimal quantum theory - quantum theory with no physical ``collapse'' of quantum states into classicality - as an ``ultimate theory that needs no modifications to account for the emergence of the classical'' (\cite{zurek:07}, p. 1).  By defining ``objectivity'' explicitly \textit{within} minimal quantum theory, it provides a philosophical basis for the ``environment as witness'' formulation of decoherence theory \citep{zurek:04, zurek:05, zurek:07} and the ``quantum Darwinist'' account of the emergence of classicality from the quantum world \citep{zurek:03, zurek:06, zurek:09}.  

By treating the environment as an all-pervasive witness that mediates and hence enables observations, the environment as witness formulation and quantum Darwinism replace the traditional concept of decoherence as an irreversible loss of quantum information \citep{zeh:70, zeh:73, zurek:81, zurek:82, joos-zeh:85, zeh:06} with a more positive concept of decoherence as an observer-independent generator of classical information.  This positive view of decoherence as both agent and explanation of the emergence of classicality has been widely adopted: classical information is now routinely defined as quantum information that has survived decoherence (e.g. \cite{griffiths:07}), decoherence-based reasoning is commonly employed to explain the post-inflation emergence of a classical universe with determinate particle masses and other properties (e.g. \cite{martineau:06, tegmark:10, kiefer:11, susskind:11}), and decoherence is taken to differentiate and hence define both Everett branches \citep{tegmark:10, wallace:10} and consistent histories \citep{griffiths:02, hartle:08, griffiths:11} in interpretations of quantum measurement (see also \cite{zurek:98, schloss:04, bacci:07, landsman:07, schloss:07, wallace:08} for general discussions of the foundational role of decoherence in the emergence of classicality).

The OPZ definition of objectivity seems straightforward: given a physical system $\mathbf{S}$, the objective properties of $\mathbf{S}$ are the properties that unbiased observers can agree about.  The present paper argues that this seeming straightforwardness is deceptive; in particular, it argues that no properties that are specified \textit{entirely within the quantum formalism} satisfy the OPZ definition of objectivity.  The only role of the OPZ definition of objectivity within decoherence theory appears, indeed, to be that of importing \textit{a priori} classical assumptions into the theory, assumptions that observers must share in order to ``arrive at a consensus'' about what they are observing and hence that violate the provisions of the OPZ definition itself that forbid ``prior agreement'' and ``prior knowledge'' on the part of the observers.

To begin the argument, it is necessary to explicitly recognize that if ``objectivity'' is to be a theoretical concept \textit{defined within} minimal quantum theory, which it must be if minimal quantum theory to is to be an ``ultimate theory that needs \textit{no modifications} - and in particular, no extra-theoretical conceptual additions - to account for the emergence of the classical,'' then observers cannot \textit{assume} that either properties or the systems they characterize are objective \textit{a priori}.  Instead, they must be able to employ the completely quantum-theoretic OPZ definition to \textit{discover} what properties of any physical system $\mathbf{S}$ are objective.  The definition must, in other words, be operational, and must apply to each quantum system for which objective properties are discoverable, as a matter of fact, by observers.  What is shown below is that the OPZ definition cannot be operationalized by observers unless they assume, \textit{a priori} and from outside of quantum theory, the very sense of ``objectivity'' that is purportedly being defined.  The paper concludes that the environment does not serve as a ``witness'' in any quantum-theoretically meaningful sense, and that the environment as witness formulation of decoherence theory does not explain the emergence of classicality.  It suggests that classicality and objectivity do not, in fact, emerge from a quantum-theoretic description of the world, but are rather imposed upon it by communities of mutually-reporting observers.

It is also necessary to state explicitly an underlying assumption: that minimal quantum theory applies universally.  This universality assumption is clearly required if minimal quantum theory is to be an ``ultimate theory that needs no modifications to account for the emergence of the classical'' and is taken to be uncontroversial in the current context.  The assumption that minimal quantum theory applies universally has two immediate consequences.  First, all \textit{physical} systems are quantum systems, and all \textit{physical} states are quantum states; there are no \textit{intrinsically} classical physical systems, and hence no systems for which classical objectivity can fairly be assumed \textit{a priori}.  Quantum theoretic specifications of systems are, therefore, taken to be ontologically definitive; any ``emergence'' of classically must be accounted for fully and completely within quantum theory.  Observers may treat the environment, each other, or their items of apparatus as classical systems, as \cite{bohr:28} insisted they must, but doing so is either an approximation or simply an error; no classical specification is ontologically definitive, and describing a system classically does not make it classical.  Second, the quantum-mechanical formalism applies, \textit{mutatis mutandis}, to \textit{all} physical systems.  It applies, in particular, to macroscopic laboratory apparatus, to the general environment, to observers, and as \cite{everett:57} insisted, to the universe as a whole.

\section{Preliminaries}

\subsection{What is a ``physical system''?}

The OPZ definition is straightforwardly about properties of physical systems: it states the conditions under which such properties may be regarded as ``objective.''  Hence the first question to ask about the OPZ definition is, what is a ``physical system''?  As the assumption that minimal quantum theory applies universally requires that any physical system be a quantum system, one might as well ask, what is a ``quantum system''?

There are clearly two kinds of answers to this question that, from a practical point of view, fall on different sides of the macroscopic - microscopic divide.  These two kinds of answers correspond to two distinct ways of saying what \textit{counts as} a physical system.  One is \textit{demonstrative}; as Fuchs puts it, ``this and this and this ... every particular that is and every way of carving up every particular that is'' (\cite{fuchs:10} p. 22).  To select a particular example, the ion-trap apparatus employed by \cite{brune:96} in their classic study of the timecourse of decoherence is a \textit{physical system}; its various parts - its vacuum chamber, ion source, pumps, magnets, lasers, detectors, read-outs and so forth - are all physical systems as well.  The second kind of answer is \textit{theoretical}.  What counts as a rubidium ion produced by the ion source in the Brune \textit{et al.} experiments is defined theoretically, as are the quantum systems comprising one or more Rb ions in interaction with the electromagnetic fields inside the ion trap.  These theoretically-defined entities are \textit{physical systems} although they can only be demonstrated by demonstrating the ion trap itself - or something like it - and invoking the theory.

The statement of the OPZ definition does not restrict it to one kind of physical system or the other; hence it is fair to assume that it is intended to apply to both.  Let `$\mathbf{S}$' name the ion trap apparatus of Brune \textit{et al.}; one can then write an instance of the OPZ definition that applies to properties of $\mathbf{S}$:

\begin{quotation}
``A property of the physical system $\mathbf{S}$ is \textit{objective} when it is:
\begin{list}{\leftmargin=2em}
\item
1. simultaneously accessible to many observers,
\item
2. who are able to find out what it is without prior knowledge about $\mathbf{S}$, and 
\item
3. who can arrive at a consensus about it without prior agreement.''
\end{list}
\end{quotation}
This specific instance of the OPZ definition specifies the conditions under which a property of the ion trap apparatus of Brune \textit{et al.} is objective, one of which is that observers ``are able to find out what (the property in question) is without prior knowledge'' about the ion-trap apparatus.  One could equally well stipulate that `$\mathbf{S}$' named the quantum system comprising whatever Rb ions were interacting with the electromagnetic fields inside the ion trap at some particular observation time $t$; in this case the OPZ definition would specify the conditions under which the properties of that system were objective, one of which is that observers would be ``able to find out what (the property in question) is without prior knowledge'' about the Rb ions or their interactions with the electromagnetic fields inside the ion trap.  

\subsection{What is a ``property'' of a physical system?}

The second question to be asked about the OPZ definition is, what is a ``property''?  What, in other words, are the \textit{candidates} for the kind of objectivity that the OPZ definition defines?  What, moreover, does it mean for a property to be a property \textit{of} a physical system?  What does it mean for a property to be a property of the ion trap apparatus of Brune \textit{et al.}, or a property of an Rb ion in interaction with an electromagnetic field?

Here again, there appear to be both demonstrative and theoretical answers.  The apparatus of Brune \textit{et al.} is (or at least was) located in Paris; ``located in Paris'' is thus a property of $\mathbf{S}$ if `$\mathbf{S}$' names the apparatus of Brune \textit{et al.}  This property is macroscopic, and it characterizes the apparatus as a bulk material object.  It is also in an important sense \textit{obvious}; it would be clear to any reasonably-aware observer, on actually encountering the apparatus, that it was located in Paris.  The principal result reported by \cite{brune:96}, on the other hand, is that controlling the interactions of the Rb ions with the electromagnetic fields inside the ion trap controls the decoherence time of the ionization state of the ions.  ``Has an ionization-state decoherence time of 38 $\mu$s'' is, therefore, a property of some Rb ions (\cite{brune:96} p. 4890).  This property is microscopic, and it characterizes a microscopic object that is only observable for a brief period of time.  It is, moreover, \textit{not} obvious; knowing what either an ionization state or a decoherence time are requires knowing at least some quantum theory.

\subsection{Properties, degrees of freedom, and Hilbert spaces}

Quantum theory characterizes physical systems in terms of physical degrees of freedom.  If $\mathbf{S}$ is a physical system, the allowed values of the degrees of freedom of $\mathbf{S}$ correspond to basis vectors of the Hilbert space describing $\mathbf{S}$.  One can, therefore, ask how the properties of a physical system that are characterized by the OPZ definition of objectivity relate to the degrees of freedom of the system, their allowed values, or the Hilbert-space representation of these allowed values.

If quantum theory applies universally to all physical systems, then it applies to the universe $\mathbf{U}$ as a whole.  One can, therefore, talk about the degrees of freedom of $\mathbf{U}$, the allowed values of these degrees of freedom, and the Hilbert space $\mathcal{H}_{\mathbf{U}}$ spanned by these allowed values.   The universal state $|\mathbf{U} \rangle$ is a vector in $\mathcal{H}_{\mathbf{U}}$; as $\mathbf{U}$ is by definition isolated, $|\mathbf{U} \rangle$ is a pure state and satisfies a Schr\"odinger equation $(\partial / \partial t) | \mathbf{U}\rangle = -(\imath / \hbar) \mathit{H}_{\mathbf{U}} | \mathbf{U}\rangle$, where $\mathit{H}_{\mathbf{U}}$ is the universal Hamiltonian.  The question of the emergence of classicality is, at bottom, a question about $\mathit{H}_{\mathbf{U}}$.  Zurek, for example, opens his classic 1998 ``rough guide'' to decoherence \citep{zurek:98} by remarking that ``it is far from clear how one can define systems given an overall Hilbert space `of everything' and the total Hamiltonian'' (p. 1794) and closes it with ``a compelling explanation of what the systems are - how to define them given, say, the overall Hamiltonian in some suitably large Hilbert space - would undoubtedly be most useful'' (p. 1818).

Any system $\mathbf{S}$ must be part of $\mathbf{U}$; hence its degrees of freedom must be among the degrees of freedom of $\mathbf{U}$.  The Hilbert space $\mathcal{H}_{\mathbf{S}}$ of any system can, therefore, be regarded as a component of a tensor-product decomposition $\mathcal{H}_{\mathbf{S}} \otimes \mathcal{H}_{\mathbf{E}} = \mathcal{H}_{\mathbf{U}}$ of $\mathcal{H}_{\mathbf{U}}$, where $\mathcal{H}_{\mathbf{E}}$ is the Hilbert space of a second system $\mathbf{E}$ conventionally called the ``environment'' of $\mathbf{S}$.  As $\mathbf{U}$ is unique, any given system $\mathbf{S}$ has a unique environment $\mathbf{E}$; the tensor-product decomposition $\mathcal{H}_{\mathbf{S}} \otimes \mathcal{H}_{\mathbf{E}} = \mathcal{H}_{\mathbf{U}}$ can thus be taken to \textit{define} $\mathbf{S}$ in terms of $\mathbf{E}$ and vice-versa.  Conversely, if no $\mathbf{E}$ exists such that $\mathcal{H}_{\mathbf{S}} \otimes \mathcal{H}_{\mathbf{E}} = \mathcal{H}_{\mathbf{U}}$, $\mathbf{S}$ cannot be a physical system.  This conception of physical systems as \textit{defined by} Hilbert-space decompositions of $\mathbf{U}$ is central to decoherence theory and to the project of explaining the emergence of classicality.  Indeed, Zurek introduces as ``axiom(o)'' of quantum theory that ``the Universe consists of systems'' and ``a composite system can be described by a tensor product of the Hilbert spaces of the constituent systems'' (\cite{zurek:03} p. 746; see also \cite{zurek:07}, p. 3; \cite{zurek:05env}, p. 2).  

With this Hilbert-space understanding of systems, it is clear that the properties referred to in the OPZ definition of objectivity can only be observable and hence allowed values of degrees of freedom.  It is, moreover, clear that the properties referred to in the OPZ definition of objectivity must be observable and hence allowed values of degrees of freedom \textit{of} particular systems.  The allowed values of position represented by basis vectors of $\mathcal{H}_{\mathbf{S}}$, for example, are allowed values of the position \textit{of} $\mathbf{S}$, or of the positions of components of $\mathbf{S}$.  Alternatively, they are allowed values of the positions of components of $\mathbf{U}$ that happen to be included within the (Hilbert-space) boundary of $\mathbf{S}$.  They are, specifically, \textit{not} allowed values of the position of $\mathbf{E}$, or of anything contained within $\mathbf{E}$, although the position of $\mathbf{E}$ may have allowed values that are numerically equal to those of $\mathbf{S}$.  It is this division of the degrees of freedom of $\mathbf{U}$, i.e. of all degrees of freedom, into degrees of freedom \textit{of} $\mathbf{S}$ and degrees of freedom \textit{of} $\mathbf{E}$ that is indicated by $\mathcal{H}_{\mathbf{S}} \otimes \mathcal{H}_{\mathbf{E}} = \mathcal{H}_{\mathbf{U}}$.  Properties of $\mathbf{S}$ and properties of $\mathbf{E}$ are similarly divided, even though as measured values they may be numerically identical.

Two opposing points of view can be taken about the degrees of freedom that characterize physical systems and the properties that correspond to their allowed values.  One point of view, exemplified by the radical anti-reductionism of \cite{fuchs:10}, treats all possible degrees of freedom on an absolutely equal basis; the objective division of the ``real world'' into systems of fixed and finite Hilbert-space dimension is taken for granted and no attempt is made to decompose any system into smaller ``building blocks'' or to reconstruct any degree of freedom from a combination of more ``fundamental'' degrees of freedom.  As Fuchs points out, the question of the emergence of classicality does not arise from this point of view; classical objectivity is simply assumed \textit{a priori}.  Indeed Fuchs ``declares the quantum-to-classical research program unnecessary (and actually obstructive)'' and insists that  ``the thing that needs insight is not the quantum-to-classical transition, but the classical-to-quantum'' (\cite{fuchs:10} p. 24, main text and fn. 46).  The alternative point of view is that embraced by the majority of working physicists.  Macroscopic systems are viewed as composites of microscopic systems from the molecular scale down to the scale of Standard Model particles, and macroscopic degrees of freedom are viewed as ``bulk'' degrees of freedom - such as total mass or center-of-mass position - that are exact mathematical consequences of the more fundamental degrees of freedom of the constituents, again down to the scale of Standard Model degrees of freedom.  From this perspective, only the most fundamental degrees of freedom need be considered when specifying the ``real'' Hilbert-space representation of any system; all other degrees of freedom ``emerge'' as physical consequences of the action of $\mathit{H}_{\mathbf{U}}$.  Indeed from this perspective, the goal of science is to explain precisely \textit{how} all non-fundamental degrees of freedom emerge as consequences of the action of $\mathit{H}_{\mathbf{U}}$.  Standard Big-Bang cosmology, with its implied hierarchy of ``special sciences'' to deal with such emergent properties as life and cognition, clearly reflects this latter perspective.   

\subsection{Observers}

The final question that must be asked about the OPZ definition of objectively is, what is an ``observer''?   As is traditional in physics, Zurek says very little about observers: ``decoherence treats observer as any other macroscopic system.  There is, however, one feature distinguishing observers from the rest of the Universe ... (they) can readily consult the content of their memory'' (\cite{zurek:03} p. 759), and later ``the observer's mind (that verifies, finds out, etc.) constitutes a primitive notion which is prior to that of scientific reality'' (p. 763-764).  Zurek does not elaborate on this characterization of observers, and does not explicitly consider whether these assumptions about observers impact the status of quantum theory as an ``ultimate theory'' or the project of explaining the emergence of classicality quantum-theoretically.  Zurek does, however, explicitly characterize the typical \textit{position} of an observer with respect to an observed system.  Even when observing a macroscopic system, Zurek points out, an observer is typically interacting with the \textit{environment} of the system - the surrounding ambient photon field, for example.  It is, indeed, this distance between the observer and the observed system that allows the environment to act as a witness, and that enables the redundant environmental encoding of information about the system that many observers can separately access \citep{zurek:04, zurek:05, zurek:07, zurek:06, zurek:09}.

This notion of a distant observer can be made precise as follows \citep{zurek:06, zurek:09}.  Let $\mathbf{S}$ be a system and $\mathbf{E}$ be its environment, i.e. let $\mathcal{H}_{\mathbf{S}} \otimes \mathcal{H}_{\mathbf{E}} = \mathcal{H}_{\mathbf{U}}$.  Let $\lbrace \mathbf{F}_{\mathit{k}} \rbrace$ be a finite set of disjoint ``fragments'' of $\mathbf{E}$, all of which are sufficiently distant from $\mathbf{S}$ and from each other to be regarded as separable from $\mathbf{S}$ and from each other for all practical purposes (FAPP).  In practice, the fragments $\mathbf{F}_{\mathit{k}}$ are assumed to be small with respect to $\mathbf{U}$; this assumption assures that $\mathbf{U}$ has sufficient degrees of freedom to fully decohere any embedded system $\mathbf{S}$ while preserving the FAPP separability of $\mathbf{S}$ from the $\mathbf{F}_{\mathit{k}}$.  Let $\lbrace \mathbf{O}_{\mathit{k}} \rbrace$ be a set of observers, $\mathbf{F}_{\mathit{k}}$ be the fragment occupied by $\mathbf{O}_{\mathit{k}}$ when $\mathbf{O}_{\mathit{k}}$ is distant from $\mathbf{S}$, and $\mathbf{E}_{\mathit{k}}$ be the environment ``shared'' by $\mathbf{S}$ and $\mathbf{O}_{\mathit{k}}$, i.e. $\mathcal{H}_{\mathbf{S}} \otimes \mathcal{H}_{\mathbf{E}_{\mathit{k}}} \otimes \mathcal{H}_{\mathbf{O}_{\mathit{k}}} = \mathcal{H}_{\mathbf{U}}$.  The fragments $\mathbf{F}_{\mathit{j}}$ occupied by all observers, and indeed all observers $\mathbf{O}_{\mathit{j}}$ with $j \neq k$, are clearly contained within the shared environment $\mathbf{E}_{\mathit{k}}$ of the $k^{th}$ observer, for all $k$.

This characterization of the positions of observers introduces a subtle but important assumption.  Consider any two observers $\mathbf{O}_{\mathit{j}}$ and $\mathbf{O}_{\mathit{k}}$, who are embedded in their respective fragments $\mathbf{F}_{\mathit{j}}$ and $\mathbf{F}_{\mathit{k}}$.  If this situation is to be well-defined, the Hilbert-space decompositions $\mathcal{H}_{\mathbf{S}} \otimes \mathcal{H}_{\mathbf{E}_{\mathit{j}}} \otimes \mathcal{H}_{\mathbf{O}_{\mathit{j}}} = \mathcal{H}_{\mathbf{S}} \otimes \mathcal{H}_{\mathbf{E}_{\mathit{k}}} \otimes \mathcal{H}_{\mathbf{O}_{\mathit{k}}} = \mathcal{H}_{\mathbf{U}}$.  In this case, however, the \textit{physics} being well defined - and in particular, the Hamiltonian $\mathit{H}_{\mathbf{U}}$ being well-defined - requires that $\mathit{H}_{\mathbf{U}}$ be independent of Hilbert-space decomposition.  This independence, called ``decompositional equivalence'' in \cite{fields:12a, fields:12b, fields:12d}, is assured by the linearity of the Hamiltonian, i.e. by $\mathit{H}_{\mathbf{U}} = \mathit{\sum_{ij} H_{ij}}$ where the indices $i$ and $j$ range over the degrees of freedom of $\mathbf{U}$, provided that \textit{only fundamental} degrees of freedom of $\mathbf{U}$ are considered, or alternatively, provided that \textit{all possible} composite degrees of freedom of $\mathbf{U}$ are considered in addition to the fundamental degrees of freedom.  What decompositional equivalence specifically disallows is that interactions between composite degrees of freedom defined with respect to some particular tensor-product decompositions of $\mathcal{H}_{\mathbf{U}}$ are included in $\mathit{H}_{\mathbf{U}}$ while others are not.  Were this to be the case, the physics of $\mathbf{U}$ as a whole would depend on its tensor-product decomposition, i.e. it would not be well-defined.  In particular, any alteration of the $\mathbf{F}_{\mathit{k}}$ or the $\mathbf{E}_{\mathit{k}}$ could, in principle, change the global physics of $\mathbf{U}$.  While it is, of course, possible that the physics of $\mathbf{U}$ as a whole is not well-defined, this possibility is inconsistent with the assumption that minimal quantum theory applies universally.

The characterization of observers as confined to distinct, separable environmental fragments also raises a question about the meaning of ``consensus'' in the OPZ definition.  The most natural interpretation of this term, in context, is the idea that the various observers make their observations, and then discuss the results to determine whether they have observed the same property $P$ of the quantum system $\mathbf{S}$.  It may be objected, however, that this is too ``human'' an interpretation, that perhaps the observers are just computers that record outcomes, as if often the case in practice.  What does ``consensus"  mean then?  If the observers employ comparable data structures and are connected by a network that supports the exchange of outcome records, ``consensus'' can clearly be achieved algorithmically.  It may, however, be the case that the observers are not so connected, and that their outcome records are collected and examined by some third party.  In this case, clearly, the relevant third party must determine whether each observer detected $P$ as a property of $\mathbf{S}$.  This is exactly the problem faced by the observers themselves in either of the previous two scenarios; hence the ``human'' interpretation can be assumed, without loss of generality, in assessing the informational requirements of achieving the required consensus.

\subsection{Einselection and the environment as witness}

Suppose $\mathbf{S}$ and $\mathbf{E}$ interact via a Hamiltonian $H_{\mathbf{S-E}}$.  The environment as witness formulation is based on two observations.  First, provided the interaction $H_{\mathbf{S-E}}$ is sufficiently larger FAPP than the self-interactions of either $\mathbf{S}$ or $\mathbf{E}$, it forces both $\mathbf{S}$ and $\mathbf{E}$ into (at least approximate) eigenstates of $H_{\mathbf{S-E}}$.  \cite{zurek:81, zurek:82, zurek:98} termed this forcing of interacting systems into eigenstates of their interaction ``environmentally-induced superselection'' or \textit{einselection}; Landsman has called einselection ``the most important and powerful idea in quantum theory since entanglement'' (\cite{landsman:07} p. 511).  Second, observers typically interact with quantum systems indirectly, by interacting with their local environments as discussed above.  An observer interacting with a local (to the observer) fragment of the environment extracts information not from the system being observed, but from an \textit{encoding} of the state of that system present in the observer's local environmental fragment.  This local fragment serves as a ``witness'' to the $\mathbf{S-E}$ interaction and hence to the state $|\mathbf{S}\rangle$ because its state $|\mathbf{F}_{\mathit{k}}\rangle$ is einselected by $H_{\mathbf{S-E}}$ \cite{zurek:04, zurek:05}; observers obtain information by querying such witnesses.  Quantum Darwinism adds to the environment as witness formulation the observation that if the fragments $\mathbf{F}_{\mathit{k}}$ are FAPP separable both from $\mathbf{S}$ and each other, multiple observers $\mathbf{O}_{\mathit{k}}$ can obtain information from their local encodings without interfering either with each other or with $\mathbf{S}$.  If this is the case, agreement among observers is possible and both $\mathbf{S}$ and its observed properties can be regarded as ``objective'' under the OPZ definition \citep{zurek:06, zurek:07, zurek:09}.  

Applications of the OPZ definition to practical measurement situations clearly require certain additional assumptions.  First, the multiple observers involved must each recognize the others as observers; otherwise the requirement for ``consensus'' cannot be fulfilled.  As discussed above, this ``recognition'' can be implemented algorithmically; it merely requires interpretability of data structures across observers, or between all the observers and some third party.  Such interpretability requires, however, that the objectivity of \textit{some} systems - the observers themselves - be assumed \textit{a priori}.  Second, the observers must agree to interact with their local fragments of the environment in comparable bases; all must, for example, agree to probe ambient visual-spectrum photons when making their observations.  This is a ``prior agreement'' among the observers, one that concerns their interactions with $\mathbf{E}$, not $\mathbf{S}$.  It introduces, however, a second objectivity assumption: the observers must jointly assume the objectivity of the physical medium - the ambient photon field, for example - that their observations probe.  Finally, the observers must share a means of communication, and this means of communication must include some symbol that is jointly agreed, or which some third party interprets, to refer to the system about which the observational consensus is to be reached.  Call this symbol `$\mathbf{S}$'.  This is also a ``prior agreement'' among the observers, or between the observers and the third party, one that concerns neither $\mathbf{E}$ nor $\mathbf{S}$ but rather the observers themselves.  As will be shown, it is the need for this final, semantic prior agreement that renders the OPZ definition inapplicable within quantum theory alone.

\section{The semantics of `$\mathbf{S}$'}
\label{Semantics}

Suppose that observers $\mathbf{O}_{\mathrm{1}}$ and $\mathbf{O}_{\mathrm{2}}$ make observations while within their respective environmental fragments $\mathbf{F}_{\mathrm{1}}$ and $\mathbf{F}_{\mathrm{2}}$, after which they confer to compare results.  Suppose further that both observers employ the name `$\mathbf{S}$' when referring to the system that they have observed, i.e. the system whose states are encoded by the states $|\mathbf{F}_{\mathrm{1}}\rangle$ and $|\mathbf{F}_{\mathrm{2}}\rangle$ with which they have respectively interacted.  One can ask how the name `$\mathbf{S}$' is being taken to refer, and whether $\mathbf{O}_{\mathrm{1}}$ and $\mathbf{O}_{\mathrm{2}}$ can provide evidence that their uses of `$\mathbf{S}$' in fact refer to the same thing.  These are not merely metaphysical questions: they are often asked seriously in practice, for example when interviewing witnesses to crimes or determining whether two graduate students are reporting observations of the same item of apparatus.  Such questions also arise in a considerably more subtle form when it becomes necessary to confirm that the degrees of freedom of an experimental apparatus have not changed - that is, when it becomes necessary to confirm that one ``system'' has not unexpectedly been replaced by a different ``system'' that may exhibit different behavior - from one replicating run of an experiment to the next.  This kind of question is regularly gotten wrong in the course of experimental research, sometimes with significant practical consequences.

\subsection{Option 1: `$\mathbf{S}$' is a rigid designator}
\label{Rigid designation}

Suppose that two observers $\mathbf{O}_{\mathrm{1}}$ and $\mathbf{O}_{\mathrm{2}}$ are observing a macroscopic system $\mathbf{S}$ that has been ``given'' demonstratively, such as the ion trap apparatus of Brune \textit{et al.}  In this case, the observers are able to directly perceive local environmental encodings of the state of $\mathbf{S}$, e.g. by interacting with the ambient photon field.  Suppose $\mathbf{O}_{\mathrm{1}}$ and $\mathbf{O}_{\mathrm{2}}$ employ the name `$\mathbf{S}$' to refer to $\mathbf{S}$, and suppose further that `$\mathbf{S}$' is a rigid designator, that is, that the name `$\mathbf{S}$' refers to $\mathbf{S}$ independently of any beliefs that employers of `$\mathbf{S}$' may have about $\mathbf{S}$, and that in fact all beliefs that some employers of `$\mathbf{S}$' have about $\mathbf{S}$ may be wrong without this affecting their reference to $\mathbf{S}$ by the use of `$\mathbf{S}$'.

Suppose now that $\mathbf{O}_{\mathrm{1}}$ and $\mathbf{O}_{\mathrm{2}}$ have positioned themselves within disjoint, effectively separable environmental fragments $\mathbf{F}_{\mathrm{1}}$ and $\mathbf{F}_{\mathrm{2}}$, have agreed to interact with their respective fragments in comparable bases - e.g. to make observations visually - and wish to employ the OPZ definition operationally to determine whether $\mathbf{S}$, that is, the system that they rigidly designate with the name `$\mathbf{S}$', has objective properties.  The OPZ definition tells them that a property $P$ of $\mathbf{S}$ is objective if it is simultaneously accessible to $\mathbf{O}_{\mathrm{1}}$ and $\mathbf{O}_{\mathrm{2}}$, if they can characterize it without knowing anything about $\mathbf{S}$, and if they can reach a consensus that each other's characterizations are correct without prior agreements about $\mathbf{S}$.  As a specific case, suppose that  $\mathbf{O}_{\mathrm{1}}$ and $\mathbf{O}_{\mathrm{2}}$ both report the property $P$ as ``indicates the pointer value 5'' after interacting with $\mathbf{F}_{\mathrm{1}}$ and $\mathbf{F}_{\mathrm{2}}$ respectively.  Are they then justified, via the OPZ definition, in concluding that $P$ is an objective property of $\mathbf{S}$?

To answer this question, it is essential to distinguish two \textit{prima facie} possibilities that the wording of the third clause, ``who can arrive at a consensus about it without prior agreement'' of the OPZ definition potentially conflates.  The most straightforward is that the ``consensus'' to be reached is that the observed property $P$ is a property of $\mathbf{S}$, the physical system rigidly designated by `$\mathbf{S}$'.  The second, less straightforward possibility is that the consensus to be reached \textit{merely concerns} $P$, irrespective of the system or systems of which $P$ is a property.  The potential conflation of these two possibilities raises a question about the OPZ definition itself: does it require that the observers also reach an evidence-based consensus - or alternatively assume \textit{a priori} - that the properties that they report are properties of a single particular system $\mathbf{S}$?  In the particular case being considered here, does the OPZ definition require that if the chosen name `$\mathbf{S}$' is a rigid designator, that the observers \textit{agree} that it is a rigid designator, and agree that their observations concern whatever `$\mathbf{S}$' designates?

This question of potential conflation is, fortunately, resolved by the formalism itself.  The environment as witness framework collapses unless the state $|\mathbf{S} \rangle$ of $\mathbf{S}$ is encoded \textit{redundantly} in multiple environmental fragments $\mathbf{F}_{\mathit{k}}$, and much recent work on quantum Darwinism has been devoted to demonstrating that redundant encoding in fact occurs \citep{zwolak:09, riedel:10, zwolak:10}.  The reason for this is obvious: two observational reports of pointer values of 5, for example, convey no information of practical use unless they also specify \textit{which pointer} was observed.  Imagine two graduate students reporting values of 5, for example, without being able to say what gauge they were reading or even whether they were reading the same one!  Without actual, in-fact encoding redundancy and hence the ability to attribute observed properties to specific systems, whether observers agree about observed properties becomes irrelevant and the notion of ``objectivity'' loses all meaning.  Redundant encoding is, however, necessary but not sufficient for objectivity under the OPZ definition: the observers must also reach a consensus that they have each, in fact, interacted with such a redundant encoding.  Thus in the above scenario, $\mathbf{O}_{\mathrm{1}}$ and $\mathbf{O}_{\mathrm{2}}$ still must establish, to each other's satisfaction, that their reports of a pointer value of 5 are due to interactions with redundant encodings of the state of $\mathbf{S}$, not to interactions with encodings of the states of systems other than $\mathbf{S}$, if they are to apply the OPZ definition.  It is worth noting that any third party receiving outcome reports from $\mathbf{O}_{\mathrm{1}}$ and $\mathbf{O}_{\mathrm{2}}$ must similarly establish the physical provenance of these reports in order to meaningfully compare them.

It is at this point that the semantics of `$\mathbf{S}$' become important.  If `$\mathbf{S}$' is a rigid designator, the reference relation between `$\mathbf{S}$' and $\mathbf{S}$ cannot depend on the beliefs of $\mathbf{O}_{\mathrm{1}}$ and $\mathbf{O}_{\mathrm{2}}$, and in particular cannot depend on their joint belief that they have observed $P$.  Hence `the thing that has $P$' - which is clearly a definite description - cannot be employed as a synonym for `$\mathbf{S}$'.  Instead, $\mathbf{O}_{\mathrm{1}}$ and $\mathbf{O}_{\mathrm{2}}$ must be able to demonstrate to each other that their observations of $P$ are due to interactions with redundant encodings of the state of $\mathbf{S}$, whatever $\mathbf{S}$ is.  That is, $\mathbf{O}_{\mathrm{1}}$ and $\mathbf{O}_{\mathrm{2}}$ must each be able to demonstrate an encoding $|\mathbf{F}_{\mathit{k}}\rangle$ in their own $\mathbf{F}_{\mathit{k}}$ with which they have interacted, and demonstrate that this $|\mathbf{F}_{\mathit{k}}\rangle$ is in fact a redundant encoding of the state of $\mathbf{S}$, whatever $\mathbf{S}$ is.  

It is clear, however, that $\mathbf{O}_{\mathrm{1}}$ and $\mathbf{O}_{\mathrm{2}}$ can make neither of these required demonstrations within the restrictions imposed by the environment as witness framework.  The environmental fragments $\mathbf{F}_{\mathrm{1}}$ and $\mathbf{F}_{\mathrm{2}}$ to which $\mathbf{O}_{\mathrm{1}}$ and $\mathbf{O}_{\mathrm{2}}$ are confined are not only disjoint but are by assumption causally decoupled FAPP; hence $\mathbf{O}_{\mathrm{1}}$ has no observational access to events occurring within $\mathbf{F}_{\mathrm{2}}$ and vice-versa.  The encodings with which $\mathbf{O}_{\mathrm{1}}$ and $\mathbf{O}_{\mathrm{2}}$ interact do not, therefore, meet the requirement of simultaneous accessibility imposed by the OPZ definition and hence cannot be regarded as objective on its criteria.  Moreover, as shown in more detail previously \citep{fields:10, fields:11}, neither $\mathbf{O}_{\mathrm{1}}$ nor $\mathbf{O}_{\mathrm{2}}$ can step out of their respective fragments and manipulate $\mathbf{S}$ directly without destroying the very encodings that they are attempting to relate to $\mathbf{S}$.  Without an ability to demonstrate that they are interacting with redundant encodings of the state of $\mathbf{S}$, however, $\mathbf{O}_{\mathrm{1}}$ and $\mathbf{O}_{\mathrm{2}}$ can employ the OPZ definition of objectivity only if they \textit{assume} that they are interacting with redundant encodings, and hence \textit{assume} that their reports of $P$ refer to properties of one particular system $\mathbf{S}$.  This is a \textit{physical} assumption about the interactions between $\mathbf{S}$, $\mathbf{E}$ and the various $\mathbf{O}_{\mathit{k}}$ that is ruled out on any reasonable reading of the OPZ definition's restrictions on ``prior knowledge'' and ``prior agreement'' among observers.  In particular, it is an assumption that is justified only if the observed property $P$ is assumed to be objective in the classical sense that the OPZ definition is attempting to define.  Hence if `$\mathbf{S}$' is regarded as a rigid designator, the OPZ definition of objectivity is impossible to apply operationally without assuming what it is attempting to define - classical objectivity - from outside of quantum theory.

\subsection{Interlude: Wigner's friend}
\label{Wigner's friend}

The classic paradox of Wigner's friend illustrates the situation faced by multiple observers of a quantum system.  In the scenario, Wigner and his friend jointly prepare a macroscopic quantum system in some state $|\mathbf{S}(\mathit{t_{0}})\rangle$, and agree that the friend will perform some particular measurement on the time evolved state $|\mathbf{S}(\mathit{t_{1}})\rangle$ at some specifed $t_{1} > t_{0}$.  Wigner then leaves the room, and returns at $t_{2} > t_{1}$ after his friend has conducted the measurement.  The paradox arises when we compare the beliefs of Wigner and his friend concerning $|\mathbf{S}(\mathit{t_{1}})\rangle$.  Having made the measurement, the friend believes that $|\mathbf{S}(\mathit{t_{1}})\rangle$ is an eigenstate of the measurement interaction.  Not having made the measurement, Wigner can only represent an entangled state $|(\mathbf{S} \otimes \mathbf{friend})(\mathit{t_{1}})\rangle$, a state that ``collapses'' only when he asks his friend about the result.

Wigner's friend is traditionally presented as an illustration of the mystery of wave-function collapse; indeed Wigner himself inferred from it that a ``being with consciousness must have a different role in quantum mechanics than the inanimate measuring device'' (\cite{wigner:62}, p. 294).  This ``mystery'' is nicely dispatched by decoherence considerations.  If one considers the environment to have witnessed and encoded every step in the scenario, Wigner can infer before re-entering the room that any $\mathbf{S} \otimes \mathbf{friend}$ entanglement was very short-lived, and that while he is ignorant of the result, his friend is not.  This straightforward and natural solution to the puzzle, however, glosses over the second deep issue raised by Wigner's friend.  By explicitly separating the joint preparation of the system from the friend's measurement, and the friend's measurement from Wigner's later query, the Wigner's friend scenario raises the question of how both Wigner and his friend \textit{re-identify} the system after preparing it.  How do they know what `$\mathbf{S}$' refers to from time to time, and how do they know that their uses of `$\mathbf{S}$' refer to the same thing?  More prosaically, how does the friend know that she has interacted at $t_{1}$ with the same system that she and Wigner jointly prepared at $t_{0}$?  If ``systems'' are collections of physical degrees of freedom, how does the friend know that she is interacting with the same collection of physical degrees of freedom at $t_{1}$ as at $t_{0}$, as opposed to interacting with more, less, or just different degrees of freedom?  These are precisely the questions of system identification and term reference with which the present paper is concerned.

As shown above, if `$\mathbf{S}$' is regarded as a rigid designator, Wigner and his friend have no way of demonstrating that the environmental encodings with which they interact are environmental encodings of $|\mathbf{S}\rangle$; even to prepare $\mathbf{S}$, they have to \textit{assume} that they are seeing and manipulating the same system.  The strength of this assumption, moreover, increases with time: when Wigner returns to the laboratory at $t_{2}$, he has to assume not only that `$\mathbf{S}$' still refers to something in the vicinity, but also that he can correctly re-identify his friend, the physical system to whom he must address his question about the results of the measurement carried out at $t_{1}$.  Note that Wigner must assume that he can correctly identify his friend even if his ``friend'' is a computer that records the results of the measurement performed at $t_{1}$, and must assume that he can correctly identify his friend even if his ``friend'' is his own memory of \textit{himself} performing the measurement at $t_{1}$.  However we are to understand these assumptions, it is clear that they are outside of quantum theory: if `$\mathbf{S}$' and other names of physical systems are regarded as rigid designators, Zurek's dream of quantum theory as an ``ultimate theory that needs no modifications to account for the emergence of the classical'' is unrealizable.

There is, however, no compelling argument that names of physical systems, at least as they are employed in the quantum-mechanical formalism, refer as rigid designators.  As noted earlier, physical systems can also be specified theoretically.  Theoretical specifications of systems are descriptive specifications, so system names based on theoretical specifications are not rigid designators but rather definite descriptions.  Given the ubiquity of theoretical specifications such as ``let $|\mathbf{S}\rangle = \mathit{\sum_{k} \lambda_{k} |s_{k}\rangle}$'' that appear to define systems in terms of collections of allowed values of degrees of freedom that are characterized by sets of Hilbert-space basis vectors, the claim that names such as `$\mathbf{S}$' typically refer as definite descriptions has at least \textit{prima facie} plausibility.  If we imagine that Wigner is using `$\mathbf{S}$' and ``my friend'' as definite descriptions, then when he returns to the laboratory he can use these terms to refer to whatever satisfies the descriptive criteria he has in mind.  In the language of decoherence, if the environmental encoding with which he is interacting appears by his chosen criteria to be an environmental encoding of his friend, that is good enough.

The use of descriptive information to identify physical systems is not only consonant with intuition and notational practice, it is a practical requirement of classical automata theory.  All systems continuously broadcast information into the environment.  An agent receiving this information is faced, first and foremost, with the source identification problem: the task of sorting out which signals are from which systems.  Suppose all the systems are finite-state machines; the source identification problem is then equivalent to the problem of assigning the most recently-received signal to a particular finite-state machine as its source.  That such an assignment cannot be made uniquely using signal data alone was proven by \cite{moore:56}: signals can only be assigned uniquely to sources if \textit{a priori} assumptions about source-signal relationships are brought to bear.  The requirement for \textit{a priori} assumptions to justify the assignment of signals to unique sources was used by \cite{chomsky:65} to demolish the then-dominate stimulus-response theory of language learning, and has formed the basis for computational analyses of perception at least since the work of \cite{marr:82} on computational models of early vision.  In the present context, the critical question is whether the \textit{a priori} assumptions required to identify physical systems as sources of observed signals can be framed \textit{and answered} within quantum theory, and whether the OPZ definition usefully captures both their framing and their answers. 

\subsection{Option 2: `$\mathbf{S}$' is a definite description}
\label{Definite description}

Suppose that `$\mathbf{S}$' is a definite description, that is, that the name `$\mathbf{S}$' refers to some physical system $\mathbf{S}$ in virtue of $\mathbf{S}$ instantiating some set $\lbrace Q_{i} \rbrace$ of properties taken to be criterial.  Suppose as above that two observers $\mathbf{O}_{\mathrm{1}}$ and $\mathbf{O}_{\mathrm{2}}$ have positioned themselves within disjoint, effectively separable environmental fragments $\mathbf{F}_{\mathrm{1}}$ and $\mathbf{F}_{\mathrm{2}}$, have agreed to interact with their respective fragments in a particular basis, and wish to employ the OPZ definition operationally to determine whether $\mathbf{S}$ has objective properties.  Clearly $\mathbf{O}_{\mathrm{1}}$ and $\mathbf{O}_{\mathrm{2}}$ cannot employ the OPZ definition to determine whether the properties $Q_{i}$ are objective; this would be blatantly circular.  Instead $\mathbf{O}_{\mathrm{1}}$ and $\mathbf{O}_{\mathrm{2}}$ must be imagined to be employing the OPZ definition to determine whether some property $P \notin \lbrace Q_{i} \rbrace$ is objective.  Such a situation is commonplace: $\mathbf{O}_{\mathrm{1}}$ and $\mathbf{O}_{\mathrm{2}}$ specify some system $\mathbf{S}$ in terms of a specified set of Hilbert-space basis vectors by saying, ``let $|\mathbf{S}\rangle = \mathit{\sum_{k} \lambda_{k} |s_{k}\rangle}$'' and proceed to interact with $\mathbf{S}$ using operators such as $\hat{x}$, $\hat{p}$ and $\hat{s}_{z}$ to detect and characterize properties other than ``is specified by the Hilbert-space decomposition $\mathcal{H}_{\mathbf{S}} \otimes \mathcal{H}_{\mathbf{E}} = \mathcal{H}_{\mathbf{U}}$.''

As might be expected, the difficulty with this commonplace scenario is that what must be established is that the eigenvalues of $\hat{x}$ \textit{acting on the system of interest} $\mathbf{S}$ are objective.  It is the eigenvalues of, e.g. $\hat{x}|\mathbf{S} \rangle$ that $\mathbf{O}_{\mathrm{1}}$ and $\mathbf{O}_{\mathrm{2}}$ must ``arrive at a consensus about ... without prior agreement.''  To reach such a consensus, $\mathbf{O}_{\mathrm{1}}$ and $\mathbf{O}_{\mathrm{2}}$ must be able to establish to each other's satisfaction that the eigenvalues that they report are eigenvalues \textit{of} $\mathbf{S}$.  Hence they must be able to show that the eigenvalues that they have measured are indeed eigenvalues of the system that they have mutually defined in terms of the particular basis vectors $\lbrace |s_{k} \rangle \rbrace$ that represent the allowed values of the specified degrees of freedom \textit{of} $\mathbf{S}$, not eigenvalues of something else.

It has been shown, however, that this cannot be done within the confines of quantum theory \citep{fields:11}.  If $\mathbf{O}_{\mathrm{1}}$ and $\mathbf{O}_{\mathrm{2}}$ are confined to $\mathbf{F}_{\mathrm{1}}$ and $\mathbf{F}_{\mathrm{2}}$, this is clear: nothing they can do from a distance can establish that the system they are interacting with is spanned exactly by a given microscopic basis $\lbrace |s_{k}\rangle \rbrace$, i.e. that what they are interacting with has the degrees of freedom for which the stipulated $|s_{k} \rangle$ represent the allowed values and no other degrees of freedom, and hence the Hilbert space $\mathcal{H}_{\mathbf{S}}$ and not some other Hilbert space.  Were $\mathbf{O}_{\mathrm{1}}$, for example, to step outside of $\mathbf{F}_{\mathrm{1}}$, reach into $\mathcal{H}_{\mathbf{S}}$ and manipulate the amplitudes of the $|s_{k}\rangle$ directly in order to determine whether the degrees of freedom within $\mathbf{S}$ and only these degrees of freedom were responsible for the observed eigenvalues, $\mathbf{O}_{\mathrm{2}}$'s observations would be disrupted by the manipulations, and the effects of the manipulations on the encoding of $|\mathbf{S} \rangle$ within $\mathbf{F}_{\mathrm{1}}$ would go unrecorded.  As in the case considered above, therefore, direct manipulations of the presumed system of interest cannot resolve the question of whether the observers are looking at the same thing.  Multiple observers can \textit{assume} that their measurements are being conducted on the same system, and hence \textit{assume} that their shared name `$\mathbf{S}$' is a definite description in the strict sense of picking out a unique referent, but they cannot demonstrate it empirically.  These assumptions are, moreover, \textit{physical} assumptions, assumptions about how the particular degrees of freedom that have been stipulated as being the degrees of freedom \textit{of} $\mathbf{S}$, those explicitly carved out by the Hilbert-space decomposition $\mathcal{H}_{\mathbf{S}} \otimes \mathcal{H}_{\mathbf{E}} = \mathcal{H}_{\mathbf{U}}$, interact with the rest of the world.  Hence they cannot be countenanced by the OPZ definition.  As in the case of a semantics based on rigid designators, a semantics based on definite descriptions renders the OPZ definition either inapplicable or pointless.

\subsection{Coda: Formal versus informal language}
\label{Informal language}

It is widely assumed that the mathematical formalism of quantum mechanics provides not only a means of calculating testable numerical predictions, but also a language for describing reality that is intrinsically more precise and conceptually less problematic than ordinary natural languages.  With this assumption comes the methodological notion that the sooner one can abandon discursive prose and start proving theorems, the better.  ``Shut up and calculate'' is the natural limiting case of this methodological notion.

As emphasized repeatedly by Bohr, however, experimentalists still speak with each other about macroscopic apparatus using ordinary language, supplemented by pointing or other indicative gestures when necessary.  In doing so, they assume the classical objectivity of the referents of their words and gestures, if not in any deep metaphysical sense, at least in the practical sense of assuming that they can know what each other are talking about.  The point of the above discussion is that theorists are no better off; they must assume that symbols like `$\mathbf{S}$' refer in the same way that experimentalists must assume that phrases like ``the ATLAS detector'' refer.  From the perspective of reference, the quantum mechanical formalism provides no help at all; questions about the reference of terms in the formalism are every bit as problematic as they are for corresponding terms in the natural languages that the formalism supplants.  It is, therefore, futile to turn to the formalism for an ``explanation'' of the appearance or ``emergence'' of objectivity that is not forthcoming using natural language terms like ``that system over there'' or ``the ATLAS detector.''  The next section examines why this is the case, by exploring in greater detail the assumptions being made when specifying that something is a ``physical system.''

\section{The underlying issue: Decompositional equivalence}
\label{Decompositional equivalence}

The treatment of macroscopic objects in classical physics involves a seldom-stated antinomy.  One the one hand, the macroscopic objects of ordinary experience are regarded as having boundaries that the laws of physics respect.  A cannonball, for example, is acted upon by the laws of classical dynamics as a single cohesive object; it is transported from place to place by mechanical forces without violations of the boundary between it and the rest of the universe.  On the other hand, such entities can be conceptually broken apart or combined together into smaller or larger systems, and the laws of physics equally applied to these.  The Earth can be considered as a system, for example, without worrying about the boundaries of the cannonballs that rest or move about on or above its surface.  Hence the laws of physics also appear \textit{not} to respect boundaries, but to allow them to be reconfigured at whim.  The first of these attitudes or assumptions is a form of commonsense realism about objects: the Moon is there, for example, as a bounded object even when no-one looks.  The second is an implicit statement of a symmetry: the laws of physics are assumed to be invariant with respect to the choice of decomposition into ``the system of interest'' and ``everything else.''  This is the symmetry of decompositional equivalence briefly introduced in \S 2.4 above.

Decompositional equivalence is a symmetry of \textit{fundamental} dynamics: it is the claim that writing the Hilbert space $\mathcal{H}_{\mathbf{U}}$ of the universe as $\mathcal{H}_{\mathbf{U}} = \mathcal{H}_{\mathbf{S}} \otimes \mathcal{H}_{\mathbf{E}}$ for some particular system $\mathbf{S}$ and its environment $\mathbf{E}$ has no impact on the Hamiltonian $H_{\mathbf{U}}$, i.e. that $H_{\mathbf{U}} = \mathit{\sum_{ij} H_{ij}}$ where $i$ and $j$ range without restriction over fundamental degrees of freedom of $\mathbf{U}$, or equivalently that $H_{\mathbf{U}} = H_{\mathbf{S}} + H_{\mathbf{E}} + H_{\mathbf{S-E}}$ for any choice of $\mathbf{S}$ and $\mathbf{E}$ for which $\mathcal{H}_{\mathbf{U}} = \mathcal{H}_{\mathbf{S}} \otimes \mathcal{H}_{\mathbf{E}}$.  Decompositional equivalence is what allows a distinction to be drawn between the fundamental dymanics $H_{\mathbf{U}}$ and the ``emergent'' dynamics $H_{\mathbf{S}}$, $H_{\mathbf{E}}$ and especially $H_{\mathbf{S - E}}$, i.e. the dynamics that only ``appear'' when an $\mathbf{S - E}$ boundary and hence a Hilbert-space decomposition of $\mathcal{H}_{\mathbf{U}}$ is specified.  Without decompositional equivalence, the interaction dynamics $H_{\mathbf{S - E}}$ would be not just different, but \textit{fundamental} for every choice of $\mathbf{S}$ and $\mathbf{E}$: different choices of system boundaries would yield different ``universes'' with different, decomposition-specific fundamental physical laws.  As pointed out in \S 2.4, without decompositional equivalence, the very idea of a ``universal dynamics'' $H_{\mathbf{U}}$ is ill-defined, and the assumption that quantum theory applies universally must be abandoned.

Decompositional equivalence is clearly a necessary condition for doing science as we know it: physics in particular would be impossible if shifting system boundaries even potentially altered fundamental dynamical laws.  Without an operating assumption that the laws of physics do not change when the boundary of the system of interest is shifted, particular collections of degrees of freedom cannot be selected as the focus of an experiment, and reliable experimental apparatus cannot be designed or constructed to target such selected degrees of freedom.  Following volumes in phase space, for example, achieves nothing without decompositional equivalence, for selecting a volume to follow could unpredictably alter the laws governing the physics of the flow.  The assumptions of isotropy and homogeneity that enable a scientific cosmology cannot be made if the choice of ``system'' affects how things work.  Without decompositional equivalence, any specified boundary encloses a ``vital essense'' that exerts an unpredictable influence on everything else.  Of recent theorists of quantum mechanics, Fuchs comes closest to denying decompositional equivalence; he explicitly rejects the reductionist idea of discovering the ``bricks with which nature is made'' (\cite{fuchs:10} p. 22), and insists that quantum systems are ``autonomous entities'' whose actions are ``moments of creation'' (p. 14).  Fuchs does not, however, take the final step of insisting that how the world is described affects its underlying laws; he takes locality of causation and separability of distant systems for granted, and offers quantum theory as a ``user's manual'' providing universal rules for rational decision making (p. 8).  This step back from the brink is understandable: decompositional equivalence is a requirement for the sense of objectivity on which commonsense realism about objects rests, and as mentioned earlier, commonsense realism about objects is something that Fuchs explicitly adopts.

Let us re-examine the notions of einselection and encoding on which the environment as witness formulation rests from the perspective of an explicit assumption of decompositional equivalence.  As before, let $\mathbf{S}$ be a system, $\mathbf{E}$ be its environment, and $\mathbf{F}_{\mathit{k}} \subset \mathbf{E}$ be a distant (from $\mathbf{S}$) fragment of $\mathbf{E}$ from within which an observer $\mathbf{O}_{\mathit{k}}$ makes observations.  The environment as witness framework is based on the idea that the interaction $H_{\mathbf{S-E}}$ correlates the state $| \mathbf{F}_{\mathit{k}} \mathit{(t)} \rangle$ of the environmental fragment $\mathbf{F}_{\mathit{k}}$ accessed by $\mathbf{O}_{\mathit{k}}$ at time $t$ with the state $|\mathbf{S} \mathit{(t)} \rangle$ of $\mathbf{S}$, and that this correlation can be considered an encoding of $|\mathbf{S} \mathit{(t)} \rangle$ by $| \mathbf{F}_{\mathit{k}} \mathit{(t)} \rangle$ from the point of view of $\mathbf{O}_{\mathit{k}}$, or of a theorist describing the situation.  Decompositional equivalence requires the laws of physics to be independent of the $\mathbf{S - E}$ decomposition, and in particular requires that $H_{\mathbf{S}} + H_{\mathbf{E}} + H_{\mathbf{S-E}} = H_{\mathbf{U}}$ for any $\mathbf{S}$ and $\mathbf{E}$; hence it requires that any encoding generated by $H_{\mathbf{S-E}}$ be equally viewable as an encoding generated purely by $H_{\mathbf{U}}$.  The universal interaction $H_{\mathbf{U}}$, however, also correlates $| \mathbf{F}_{\mathit{k}} \mathit{(t)} \rangle$ with the state $|\mathbf{S}^{\prime} \mathit{(t)} \rangle$ of every alternative system $\mathbf{S^{\prime}}$ that is sufficiently distant from $\mathbf{F}_{\mathit{k}}$ for FAPP separability \citep{fields:12b}.  Hence \textit{all possible} systems, including all possible systems containing, contained in, or overlapping with $\mathbf{S}$, must simultaneously generate encodings in every distant environmental fragment, including in particular $\mathbf{F}_{\mathit{k}}$.  Nothing, moreover, requires the spatial boundaries of physical systems to be simply connected; physicists study many systems - most notoriously, spatially-separated pairs of particles in asymmetric Bell states - that do not have simply connected spatial boundaries.  The requirement that a fixed set of physical laws - and hence a fixed dynamics $H_{\mathbf{U}}$ - generates encodings of \textit{all possible} physical systems therefore also applies to systems comprising degrees of freedom that are not contiguous in either position or momentum space. 

Decompositional equivalence thus requires the state $|\mathbf{S^{\prime}}\mathit(t)\rangle$ of any possible physical system $\mathbf{S^{\prime}}$ to be einselected, at every instant $t$, by its interaction with its environment $\mathbf{E^{\prime}}$, an environment that by assumption contains $\mathbf{F}_{\mathit{k}}$ as a distant and FAPP separable fragment.  One can now ask, what do the encodings in $\mathbf{F}_{\mathit{k}}$ of these einselected states $|\mathbf{S^{\prime}}\mathit(t)\rangle$ look like?  The number of possible systems $\mathbf{S^{\prime}}$ increases combinatorially with the number of degrees of freedom of $\mathbf{U}$; for any physically-reasonable $\mathbf{F}_{\mathit{k}}$, the number of possible $\mathbf{S^{\prime}}$ will be far larger than the number of degrees of freedom of $\mathbf{F}_{\mathit{k}}$, and can be regarded as arbitrarily large FAPP.  Any particular encoding $|\mathbf{F}_{\mathit{k}}\mathit(t)\rangle$ will, therefore, be ambiguous between an arbitrarily large number of einselected states $|\mathbf{S^{\prime}}\mathit(t)\rangle$, and hence ambiguous between an arbitrarily large number of encoded systems $\mathbf{S^{\prime}}$ \citep{fields:13a}.  Call the set of einselected states $|\mathbf{S^{\prime}}\mathit(t)\rangle$ that are encoded by $|\mathbf{F}_{\mathit{k}}\mathit(t)\rangle$ ``$Im^{-1}(|\mathbf{F}_{\mathit{k}}\mathit(t)\rangle)$,'' the inverse image in $\mathbf{U}$ of the einselection-driven encoding $|\mathbf{F}_{\mathit{k}}\mathit(t)\rangle$.  To an observer $\mathbf{O}_{\mathit{k}}$ restricted to $\mathbf{F}_{\mathit{k}}$, the states in $Im^{-1}(|\mathbf{F}_{\mathit{k}}\mathit(t)\rangle)$ and hence the alternative system - environment decompositions that they characterize are indistinguishable by any physical means.  Hence a property $P$ being detectable by measurement within  $\mathbf{F}_{\mathit{k}}$ conveys to an observer within $\mathbf{F}_{\mathit{k}}$ only information about $| \mathbf{F}_{\mathit{k}}\rangle$; it conveys no more information about the world outside of $\mathbf{F}_{\mathit{k}}$ than that its state is consistent with measurements of $|\mathbf{F}_{\mathit{k}}\rangle$ revealing $P$.  In particular, it conveys no other \textit{classical} information about the world outside of $\mathbf{F}_{\mathit{k}}$, despite the ubiquitous action of decoherence and einselection.  The inability of $\mathbf{O}_{\mathit{k}}$ to determine the referent of a formal symbol such as `$\mathbf{S}$' from within $\mathbf{F}_{\mathit{k}}$ is, therefore, due not to any failure of language, but to decompositional equivalence; `$\mathbf{S}$' is ambiguous as a matter of \textit{physics} between the arbitrarily-large number of alternative systems for which states are physically encoded in $\mathbf{F}_{\mathit{k}}$ by decoherence and einselection.

The assumption of decompositional equivalence, therefore, undoes the very work of classical information creation that decoherence and einselection were designed to do.  Without a specific and restricted set of decompositional boundaries at which to act, decoherence acts everywhere, creating information about all possible things and hence about nothing.  By witnessing everything all the time, the environmental fragment $\mathbf{F}_{\mathit{k}}$ encodes not easily-extracted classical information about particular systems but rather decomposition-independent quantum information about the entire rest of the universe.  Such witnesses are of no help to observers; instead of too little, they know far too much.

This understanding of decompositional equivalence allows us, finally, to confront the common notion that ``high-level'' properties of $H_{\mathbf{U}}$ somehow result in certain system-environment boundaries being ``preferred'' in some objective sense.  We can now ask: \textit{preferred by what?}  The environment as witness formulation of decoherence theory is based on the idea that the \textit{environment} prefers some systems over others, but we have seen that this cannot be the case: $H_{\mathbf{U}}$ forces any environmental fragment $\mathbf{F}_{\mathit{k}}$ to encode einselected states of huge numbers of alternative systems promiscuously, and from the perspective of any observer confined to that fragment, ambiguously.  If preferences on the part of the environment are rejected, however, the only remaining answer is \textit{preferred by us}.  It is perfectly obvious why \textit{we} might prefer some system-environment boundaries over others; as living organisms we care about what is happening at some system-environment boundaries - our own skins, for example - and not others.  That our preferences \textit{not matter} is, however, a requirement of ``objectivity'' if anything is.  We assume decompositional equivalence as a basis for doing science in order to assure that it has some possibility of being objective.  We cannot both do this, and assume that the boundaries we prefer have any impact on the action of $H_{\mathbf{U}}$.

\section{Conclusion}
\label{Conclusion}

The OPZ definition of objectivity is an attempt to characterize, in operational terms, how observers can discover a classically objective world.  It assumes that classical objectivity has a physical explanation, and hence motivates the formal construction of such an explanation, quantum Darwinism, through the use of the environment as witness formulation of decoherence theory.  The OPZ definition requires that observers have ``no prior knowledge'' of the systems that they observe, and make ``no prior agreements'' about their properties; however, it implicitly assumes that the Hilbert-space boundaries of systems of interest to observers have been specified.  It is shown here that any such specification can only come from the observers themselves; hence if the OPZ definition is not empty, it is circular.  It cannot, therefore, informatively characterize how observers behave, and it cannot effectively motivate the construction of a physical theory.

The problem with the OPZ definition is, however, not merely that it is inapplicable: it is the underlying assumption that classical objectivity emerges via an observer-independent physical mechanism.  Decoherence and einselection appear to explain classicality in an observer-independent way, but do so only if the particular systems on which they are taken to be acting are specified in advance.  \cite{zurek:03} attempts to justify such a specification by assuming the existence of systems as ``axiom(o)''; however, this assumption is not strong enough to produce a classically-objective world containing only the systems that human observers in fact observe.  Classical objectivity ``emerges'' by the mechanisms of decoherence and einselection only if ``axiom(o)'' is replaced by a much stronger axiom: ``The Universe consists of \textit{these} systems'' followed by a listing of some Hilbert-space decompositions at the expense of all others.  A universe in which such a system-specifying axiom holds is a universe in which physical dynamics is forced to respect a fixed set of pre-established system boundaries, i.e. it is a universe in which decompositional equivalence is false.

We have, however, no evidence that we live in a universe in which decompositional equivalence is false, and the evidence of successful science to argue that it is not.  We must, therefore, look elsewhere for the origin of classicality.  The most obvious option is that the origin of classicality lies not in physics but in us: in the very definitions of observation and communication.  \cite{bohr:28} argued for this option on the basis of common sense, \cite{moore:56} did so on the basis of classical automata theory, and \cite{roederer:05} does so on the basis of evolutionary biology and neuroscience.  Anyone who embraces Landsman's ``stance 1'' that ``quantum theory is fundamental and universally valid, and the classical world has only `relative' or `perspectival' existence'' (\cite{landsman:07} p. 425) should not be uncomfortable with this position.

\section*{Acknowledgement}

I thank Juan Roederer for informative and enjoyable discussions of the issues explored in this paper, and two anonymous referees for raising useful questions about an earlier version.

\end{document}